\documentclass[
    ,final            
  ]
  {aipproc}

\layoutstyle{8x11double}

\usepackage{graphicx}
\usepackage[figuresright]{rotating}

\newcommand{\bez}{\begin{eqnarray*}}
\newcommand{\eez}{\end{eqnarray*}}
\newcommand{\be}{\begin{equation}}
\newcommand{\ee}{\end{equation}}
\newcommand{\beq}{\begin{eqnarray}}
\newcommand{\eeq}{\end{eqnarray}}
\newcommand{\bc}{\begin{center}}
\newcommand{\ec}{\end{center}}

\def\msun{{\rm M_\odot}}
\def\xte{{\it RXTE}}
\def\rinf{r_{\infty}}
\def\Tinf{T_{\infty}}
\def\d{{\rm d}}
\def\rg{r_g}
\def\betaeq{\beta_{\rm eq}}
\def\Dop{\delta}
\def\taut{\tau_{\rm es}}
\def\source{{SAX J1808.4$-$3658}}
\def\degr{^{\rm o}}

\begin{document}


\title{The Physics of X-ray Emission \\
from Accreting Millisecond Pulsars}
\author{Juri Poutanen}{address={Astronomy Division, P.O. Box 3000,
        FIN-90014 University of Oulu, Finland}}        

\begin{abstract}
By analyzing the  {\it Rossi X-ray Timing Explorer} data on \source, 
we show that the  X-ray emission in accretion powered
millisecond pulsars can be produced  by  Comptonization  in a hot slab
(radiative shock) of  Thomson  optical  depth  $\taut\sim1$ at the
neutron  star surface.  The escaping radiation consists of two components:
a black body and a hard Comptonized tail. These components 
have very different angular distribution: 
the black body peaks along the slab normal (a ``pencil''--like emission pattern),
while the tail has a broader angular distribution (a ``fan''--like pattern).
This results in  very different variability properties.
We construct a detailed model of the X-ray production accounting for the
Doppler  boosting,  relativistic   aberration  and  gravitational  light
bending.
We are able to reproduce the pulse profiles at different energies, 
corresponding phase lags, as well as the  time-averaged  spectrum.
We obtain constraints on the neutron star radius: $R\sim11$
km if its mass $M=1.6\msun$, and $R\sim8.5$ km if  $M=1.4\msun$. 
We  present simple analytical formulae for  computing the
light curves and oscillation amplitudes expected from hot spots in 
X-ray bursters and accretion powered millisecond pulsars.
We also propose an analytical expression that can be used to
determine the size of the black body emission region from the
observed properties.
\end{abstract}

\maketitle


\section{INTRODUCTION}

In order to understanding the X-ray production in accreting black hole and 
neutron star sources in X-ray binaries, one needs to model their 
spectral and temporal properties simultaneously. 
In the black hole case, the emission is believed to originate in the accretion 
disk and/or its corona. Detailed spectral fits to the broad band spectra leave little 
doubts that the main emission mechanism is thermal or non-thermal 
Comptonization. However, in spite of extensive efforts, the exact geometry of the 
emission region is still not known \cite{p98}. The hard X-rays can be 
produced in the central hot flow \cite{esin98,pkr97}
and/or in the magnetically dominating 
corona/jet in the vicinity of a
cooler accretion disk \cite{b99,mal01}.
Temporal variability provide some additional constraints 
on the size of the emission region \cite{mac00}, but 
has difficulties in breaking the model degeneracy \cite{p01}. 
The observed spectral-temporal correlations seem, however, 
to favor the hot inner disk scenario  \cite{gilf99,gilfthis,aaz03}
at least for the hard state dominated by thermal Comptonization.

In the case of accreting neutron stars  in low-mass X-ray binaries,
 the emission can be
produced in the accretion disk (and corona) as well as at the neutron 
star surface, with the later dominating the total energy output
\cite{ss00,ss88}. 
Depending on the magnetic field strength and the accretion rate, 
there can be different scenarios. The gravitational energy can be
dissipated  gradually in the boundary (or spreading) layer, while the matter is
slowing down from the rapid Keplerian rotation to a slower rotation of a
neutron star, or abruptly in the shock when the accreting material,
following magnetic field lines, is channeled towards magnetic poles of the star.
Direct spectral decomposition can be quite complicated \cite{done03}.
The analysis of the Fourier frequency resolved spectra of neutron stars in
low-mass X-ray binaries shows that the accretion
disk producing soft X-ray emission does not vary much, while the 
hard radiation (from the boundary layer?) 
is responsible for most of the observed variability at 
high Fourier frequencies \cite{gilf03}. 

In all the cases considered above, it is not absolutely clear 
what is the origin of a given spectral component.  
In  accretion powered \cite{wij03} 
as well as nuclear powered (X-ray bursters) \cite{sb03}
millisecond pulsars, the situation seems to be simpler.
Observations of the coherent pulsations confirms that 
the bulk of the emission is produced at the surface of a
neutron star  which rotates at 200--600 Hz.
We can also fold the light curve on a pulsar period
and study the pulse profile as a function of energy.
For accreting pulsars the statistics is better since 
one can fold the profiles over a longer observational period
(days rather than seconds in the X-ray burst oscillations). 
Detailed modeling of the pulse profiles allows us 
to get interesting constraints on the compactness  of the neutron star
as well as on the emission  pattern from the
neutron  star  surface \cite{pg03}.

\section{SPECTRUM AND RADIATION PATTERN}

\begin{figure}
\includegraphics[width=\columnwidth]{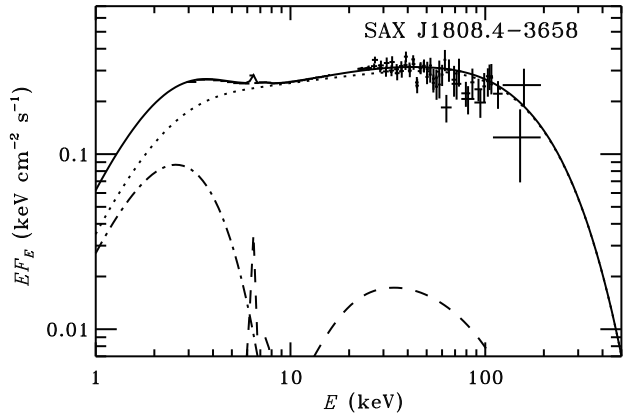}
\caption{Time-averaged spectrum of SAX J1808.4$-$3658  as observed
by \xte\  and the model spectrum. See text for details.}
\label{fig:spectr}
\end{figure}

We assume that the emission originates close to the  neutron star surface
in an accretion shock which we approximate as a plane-parallel slab.
The time-averaged observed spectrum of the best studied accretion
powered pulsar \source\ 
(see Fig.~\ref{fig:spectr}) is similar to
those in other neutron star and black hole sources.
It can be represented as a composition of a black body  with temperature
$kT_{\rm bb}=0.66$ keV (dash-dotted curve), thermal Comptonization (dots)
by electrons of temperature $\sim50-100$ keV  and optical depth $\taut\sim 1$,
Compton reflection bump and the iron line (dashes).
The Compton reflection (from the disk)
has a rather small amplitude because of the
disruption of the accretion flow in the neutron star vicinity \cite{gier}.

\begin{figure}
\includegraphics[height=.35\textheight]{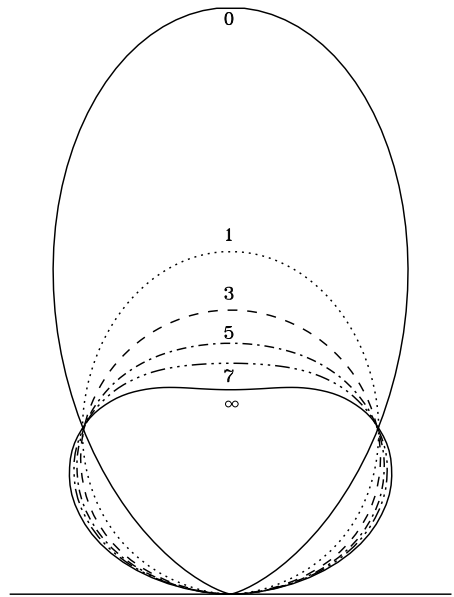}
\caption{Polar diagram of the normalized radiation flux 
 $\mu I(\mu)$ escaping from an
electron scattering slab of Thomson optical depth $\taut=0.7$.
Here  $\arccos\mu$ is the 
angle between the normal and the photon direction. 
Different scattering orders are shown and marked by numbers.
}
\label{fig:rad}
\end{figure}

If the settling bulk velocity smaller
than the thermal electron velocity, we can easily compute the angular
distribution of the escaping radiation \cite{ps96,st85}.
The radiation pattern is very different for photons of  different
scattered orders (see Fig.~\ref{fig:rad}).  One sees that
the black body (marked with 0) is strongly peaked along the
normal to the slab, while photons scattered many times are beamed in
the direction making an angle $\sim 50\degr-60\degr$ from the normal.
In the Comptonization process, scatterings also shift photons along the
energy axis so that higher energy photons are scattered more times.

\begin{figure}
\includegraphics[height=.4\textheight,width=\columnwidth]{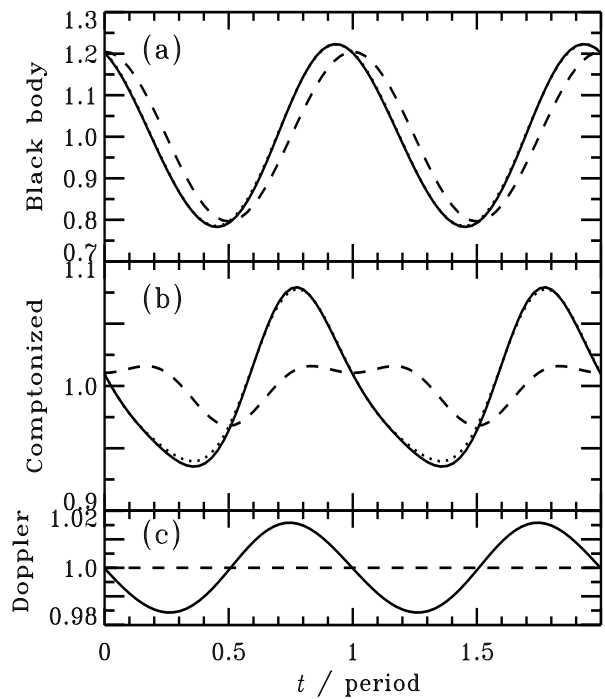}
\caption{Light curves expected from a slowly rotating star
(dashed curves) and that rotating at 401 Hz (solid curves).
Parameters: $M=1.4\msun$, $R=2\rg=8.4$ km, $i=80\degr$, $\theta=11\degr$,
$\tau=0.16$ and $a=-0.78$ (this parameter corresponds to
the scattering optical depth $\taut\sim 0.7$). From \cite{pg03}.}
\label{fig:lc}
\end{figure}

\section{MODEL AND RESULTS}

\begin{figure}
\includegraphics[width=\columnwidth]{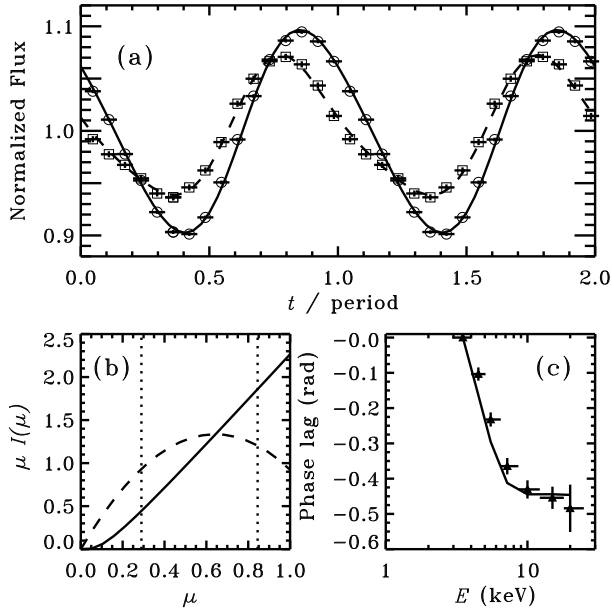}
\caption{(a) Pulse profiles of SAX J1808.4$-$3658 in the
3--4 keV (circles) and 12--18 keV (squares) energy band and
the model light curves (solid and dashed curves).
Same parameters as in Fig.~\ref{fig:lc}.
(b) The angular distribution of the intrinsic black body
(solid curve) and Comptonized (dashed curve) fluxes $\mu I(\mu)$
in the spot co-rotating frame normalized as $\int \mu I(\mu) d \mu=1$.
Only the range of angles between the dotted lines is actually observed.
(c) The observed (crosses) and the model (solid curve) phase lags
at the pulsar frequency relative to the 3--4 keV band. From \cite{pg03}.
}
\label{fig:all}
\end{figure}

The light curves from a circular spot
are computed accounting for the special  relativistic  effects  (Doppler
boosting, relativistic aberration) and the
gravitational light bending in Schwarzschild metric.
The model is described in details in \cite{pg03}.
The spot spectrum is assumed to consist
of two components: a black body and a Comptonization tail with
different angular distributions.
Their spectral shapes are taken as in the observed
time-averaged spectrum (see Fig.~\ref{fig:spectr}).
The model parameters are:
the pulsar frequency (fixed at $\nu=401$ Hz for  \source),
the neutron  star mass $M$,
stellar radius $R$, inclination $i$,
colatitude of the spot center $\theta$,
``optical  depth'' $\tau$, that describes the
angular dependence of the black body
intensity $I_{\rm bb}(\mu)\propto \exp(-\tau/\mu)$, and  a
parameter $a$ determining  the
angular distribution of the Comptonized radiation 
$I_{\rm sc}(\mu)\propto 1+a\mu$.
This linear dependence of the intensity mimics the angular distribution
of radiation escaping from a slab (Fig.~\ref{fig:rad}).

A rapid rotation of the star causes significant changes in the pulse
profile. We illustrate this by showing in Fig.~\ref{fig:lc} 
the light curves of a slowly rotating star (dashed curves) and those 
modified by the Doppler boosting and aberration for a neutron star of
rotational frequency 401 Hz (solid curves).
We consider both the black body and the Comptonized emission.
The pulse profiles  strongly depend on the assumed emission pattern.
Light bending depends on the compactness of the star $R/\rg$ (where
$\rg=2GM/c^2$) and reduces the variability amplitude with respect
to a less compact star (see below).

\begin{figure}
\includegraphics[height=0.34\textheight]{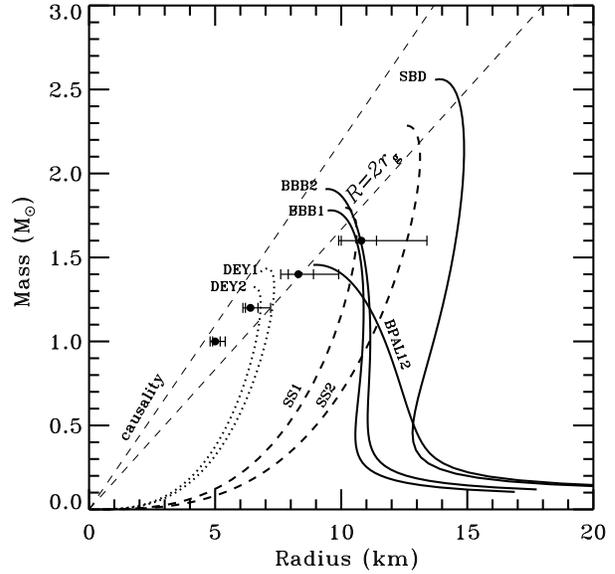}
\caption{Constraints  on the radius of the compact star from
the light curve of SAX J1808.4$-$3658 are shown by
circles with double error bars (corresponding to 90\% and 99\% confidence
limits). Different equations of state for strange and neutron stars are
shown for comparison. See \cite{pg03} for details. }
\label{fig:eos}
\end{figure}

\newpage

The model parameters can be constrained
by fitting the observed light curves from \source\ 
(see Fig.~\ref{fig:all}).
The fitted emission pattern of Comptonized and black body radiation
confirms our expectations: the black body is beamed along the surface normal
while the hard radiation flux (proportional to the 
intensity times the cosine of the projected area) is more isotropic 
(Fig.~\ref{fig:all}b). 
One of the main result of this study is
determination of the radius of the compact star.
The dependence of the radius on the assumed mass is
shown in Fig.~\ref{fig:eos}.
For a standard neutron star mass of $1.4\msun$, we get 
$R=8.5\pm0.5$ km, while for a more massive star 
$M=1.6\msun$, the radius of 11 km is consistent with some 
neutron star equation of state. 
The  inclination of the system can be constrained to $i>65\degr$.
Our model is able to fit the time-averaged spectra, the energy dependent 
pulse profiles and the observed phase lags,
simultaneously.

Very similar pulse profiles (but with higher oscillation 
amplitude) were also observed from the 
recently discovered fifth ms pulsar XTE~J1814$-$338 \cite{s03}.
In the framework of the present model, the data can be explained
for example, by increasing the colatitude of the spot center 
$\theta$ from $11\degr$ to $\sim 17\degr$. The presence of the 
harmonic in the burst oscillations \cite{s03}
could result from the larger contribution of the Comptonized 
radiation (with a ``fan''--like emission pattern) 
to the observed flux in this source comparing to other X-ray bursters
since a black body does not produce harmonics (see eq.~\ref{eq:amplani}). 
This interpretation is supported by the observed hard 
spectrum \cite{m03}.

\section{ANALYTICAL LIGHT CURVES, OSCILLATION AMPLITUDES, AND SPOT SIZE}

Using formalism described in \cite{b02} (see \cite{pg03} for details)
we obtain simple expressions for the light curve
from a small spot at a rapidly rotating neutron star. 
For example,
the bolometric flux from a black body spot  is
\be \label{eq:dF1}
\d F=  \left( 1-\frac{\rg}{R} \right)^2
\Dop^5 \left[ \frac{\rg}{R}+ \left(1-\frac{\rg}{R}
\right)  \cos\psi  \right] I_0 \frac{\d S}{D^2} ,
\ee
where $I_0$ is the black body intensity (in the spot comoving
frame), $\d S$ is the spot area, $D$ is the distance to the
source, $\cos\psi=\cos i\ \cos\theta+\sin i\ \sin \theta\ \cos\phi$,
$\phi$ is the pulse phase, and $\Dop$ is the Doppler factor.
We can also get simple (but very accurate) expression
for the oscillation amplitude (here from one spot and neglecting
Doppler boosting which does not affect it much):
$A \equiv (F_{\max}-F_{\min})/(F_{\max}+F_{\min})= U/Q$,
where we defined
$U=(1-\rg/R) \sin i\ \sin \theta$ and 
$Q=\rg/R+(1-\rg/R) \cos i \ \cos\theta$ .

The time-averaged flux observed from a spot of 
angular radius $\rho$ is \cite{pg03}
(assuming the spot is always visible)
\be
F_{\rm bb}=\left( 1-\frac{\rg}{R} \right)^2
\left[ Q + \frac{\rg}{R} \tan^2\frac{\rho}{2} \right]
I_0 \pi R^2 \sin^2\rho /D^2 .
\ee
This formula allows us to obtain
a simple expression for estimating the  
emission region size (compare to \cite{pod00}):
\be
R\ \sin\rho = \rinf \left( Q +
\frac{\rg}{R} \tan^2\frac{\rho}{2} \right)^{-1/2} ,
\ee
which can be solved by iterations. Here $\rinf$ is the
``observed" spot size,
$F_{\rm bb}=\sigma_{\rm SB} \Tinf^4 \rinf^2 /D^2$, and
$\Tinf$ is the fitted temperature.
For small $\rho$, we get
$R\rho = \rinf Q^{-1/2}$. 
For example, if we look at the spot always along the symmetry axis
(take $i=0,\ \theta=0$), we get $Q=1$ and $R\rho=\rinf$. Thus in that
case, light bending does not affect the size.
A spot of a finite size produces a smaller variability
amplitude
\be \label{eq:amplspot}
\strut\displaystyle
A=U/\left[ Q+\frac{\rg}{R} \tan^2\frac{\rho}{2} \right].
\ee

For an anisotropic source 
given by a linear relation $I(\mu)\propto 1+a\mu$ at 
a slowly rotating star ($\delta=1$), we can obtain
the ratio of the variability amplitudes at the fundamental
frequency $A_0$ and the harmonic $A_1$, using equation (\ref{eq:dF1}):
\be \label{eq:amplani}
\frac{A_1}{A_0}=\frac{aU/2}{1+2aQ}.
\ee
One sees that a black body source $a=0$ 
does not produce any harmonics.
We can estimate the impact of rapid rotation on that
ratio. Expanding Doppler factor in powers of the equatorial
velocity $\betaeq=v_{\rm eq}/c$ and keeping the first term only,
we get (see \cite{pb04} for details):
\be
\frac{A_1}{A_0}\approx \frac{5}{2} \betaeq \sqrt{1-\rg/R}\ \sin i\ \sin\theta .
\ee
Thus this ratio depends linearly on $\sin i$ and $\sin\theta$ (as 
for anisotropic source, eq.~\ref{eq:amplani}). 
One also sees that due to a linear dependence on the rotational frequency, 
it is much easier to get large amplitude of the 
harmonic with an anisotropic source than just by rotation. 
These expressions can be used
for analysis of the data 
from ms pulsars and X-ray bursters.

\begin{theacknowledgments}
This research was supported by the Academy of Finland,
the Jenny and Antti Wihuri Foundation, and 
the Nordic project in High Energy Astrophysics (NORDITA). 
\end{theacknowledgments}

\end{document}